\documentclass{emulateapj}

\usepackage{amsmath}
\usepackage{amssymb}
\usepackage{graphicx}
\usepackage{color}
\usepackage{natbib}
\usepackage{epsfig}

\def\gtaprx {\lower .1ex\hbox{\rlap{\raise .6ex\hbox{\hskip .3ex
	{\ifmmode{\scriptscriptstyle >}\else
		{$\scriptscriptstyle >$}\fi}}}
	\kern -.4ex{\ifmmode{\scriptscriptstyle \sim}\else
		{$\scriptscriptstyle\sim$}\fi}}}
\def\ltaprx {\lower .1ex\hbox{\rlap{\raise .6ex\hbox{\hskip .3ex
	{\ifmmode{\scriptscriptstyle <}\else
		{$\scriptscriptstyle <$}\fi}}}
	\kern -.4ex{\ifmmode{\scriptscriptstyle \sim}\else
		{$\scriptscriptstyle\sim$}\fi}}}

\newcommand{\cutt}[1]{\textcolor{blue}{}}

\newcommand{\Ms}{{\ensuremath{{M}_{\odot} }}}
\newcommand{\Zs}{\ensuremath{Z_\odot}}

\newcommand{\HII}{{\ion{H}{2}}}

\begin{document}

\title{The Supernova that Destroyed a Protogalaxy:  Prompt Chemical Enrichment 
and Supermassive Black Hole Growth}

\author{Daniel J. Whalen\altaffilmark{1,2},  Jarrett L. Johnson\altaffilmark{1}, 
Joseph Smidt\altaffilmark{1}, Avery Meiksin\altaffilmark{3}, Alexander 
Heger\altaffilmark{4}, Wesley Even\altaffilmark{5} and Chris L. 
Fryer\altaffilmark{5}}

\altaffiltext{1}{T-2, Los Alamos National Laboratory, Los Alamos, NM 87545}

\altaffiltext{2}{Universit\"{a}t Heidelberg, Zentrum f\"{u}r Astronomie, Institut f\"{u}r 
Theoretische Astrophysik, Albert-Ueberle-Str. 2, 69120 Heidelberg, Germany}

\altaffiltext{3}{Institute for Astronomy, University of Edinburgh, Blackford Hill, 
Edinburgh EH9 3HJ, UK}

\altaffiltext{4}{Monash Centre for Astrophysics, Monash University, Victoria, 
3800, Australia}

\altaffiltext{5}{CCS-2, Los Alamos National Laboratory, Los Alamos, NM 87545}

\begin{abstract}

The first primitive galaxies formed from accretion and mergers by $z \sim $ 15, 
and were primarily responsible for cosmological reionization and the chemical 
enrichment of the early cosmos.  But a few of these galaxies may have formed 
in the presence of strong Lyman-Werner UV fluxes that sterilized them of H$_2$, 
preventing them from forming stars or expelling heavy elements into the IGM 
prior to assembly.  At masses of 10$^8$ \Ms\ and virial temperatures of 10$^4$
K, these halos began to rapidly cool by atomic lines, perhaps forming 10$^4$ - 
10$^6$ \Ms\ Pop III stars and, later, the seeds of supermassive black holes.  We 
have modeled the explosion of a supermassive Pop III star in the dense core of 
a line-cooled protogalaxy with the ZEUS-MP code.  We find that the supernova
(SN) expands to a radius of $\sim$ 1 kpc, briefly engulfing the entire galaxy, but 
then collapses back into the potential well of the dark matter.  Fallback fully 
mixes the interior of the protogalaxy with metals, igniting a violent starburst and 
fueling the rapid growth of a massive black hole at its center. The starburst would 
populate the protogalaxy with stars in greater numbers and at higher metallicities 
than in more slowly-evolving, nearby halos.  The SN remnant becomes a strong 
synchrotron source that can be observed with eVLA and eMERLIN and has a 
unique signature that easily distinguishes it from less energetic SN remnants.  
Such explosions, and their attendant starbursts, may well have marked the 
birthplaces of supermassive black holes on the sky. 

\vspace{0.1in}

\end{abstract}


\section{Introduction}

After the appearance of the first stars ended the cosmic Dark Ages at $z \sim$ 
25 \citep{bcl99,abn00,abn02, bcl02,nu01,on07,on08,wa07,y08,turk09,stacy10,
clark11,sm11,get11,get12}, primeval galaxies formed by accretion and mergers 
between cosmological halos by $z \sim$ 15 \citep{jgb08,get08,jlj09,get10,jeon11,
pmb11,wise12,pmb12}.  Radiation and strong winds from these galaxies began 
to reionize \citep{wan04,ket04,abs06,awb07,wa08a} and chemically enrich the 
IGM \citep{mbh03,ss07,bsmith09,jet09b,jw11,ritt12,chiaki12}.  

A few of these galaxies form in the vicinity of strong sources of Lyman-Werner 
(LW) UV flux that sterilize them of H$_2$, preventing them from cooling and
forming primordial stars or hosting SN explosions prior to assembly \citep[e.g.,
][]{jlj12a}.  These protogalaxies reach masses of 10$^8$ \Ms\ and virial 
temperatures of 10$^4$ K without having blown any gas into the IGM or been 
chemically enriched. At 10$^4$ K, gas in these halos began to cool by H lines, 
triggering catastrophic baryon collapse at their centers with infall rates of up to 
1 \Ms\ yr$^{-1}$ \citep{wta08,rh09,sbh10,whb11}.  These rates are 1000 times 
those that created the first stars and may have led to the formation of 
supermassive gas fragments.  

In most cases these fragments bypassed star formation and collapsed directly 
to 10$^4$ - 10$^5$ \Ms\ black holes (BHs).  These objects may have been the 
progenitors of the supermassive black holes (SMBHs) found in most massive 
galaxies today \citep{bl03,begel06,jb07b,brmvol08,lfh09,th09,pan12b}.  The 
creation of SMBH seeds by direct baryon collapse is favored by many because 
it is difficult for the BHs of Population (Pop) III stars to sustain the rapid growth 
needed at early times to reach 10$^9$ \Ms\ by $z \sim$ 7 \citep{fan03,wm03,
fan06,milos09,awa09,pm11,mort11,pm12,wf12,pm13,jet13}.  It was originally 
believed that LW fluxes capable of fully quenching Pop III star formation were 
rare, and that this might explain why only $\sim$ one 10$^9$ \Ms\ BH is found 
per Gpc$^{-3}$ at $z \sim$ 7, but it has recently been discovered that such 
conditions may have been far more common in the early universe than 
previously thought \citep{dijkstra08,jlj12a,agarw12,petri12}.  

Some supermassive clumps in line-cooled halos formed stable stars instead 
of collapsing directly to BHs \citep{iben63,fh64,fowler66,af72a,af72b,bet84,
fuller86,begel10}.  It is now known that some of these stars died in the most 
energetic thermonuclear explosions in the universe \citep{montero12,heg13}.  
Recent simulations have shown that such SNe will be visible in both deep-field 
and all-sky NIR surveys at $z \gtrsim$ 10 by the \textit{James Webb Space 
Telescope} (\textit{JWST}) and the Wide-Field Infrared Survey Telescope 
(WFIRST) and Wide-Field Imaging Surveyor for High-Redshift (WISH) \citep{
wet12d} \citep[see also][for other work on Pop III SN light curves]{sc05,fwf10,
kasen11,pan12a,hum12,det12,wet12a,wet12b,wet12c,wet12e,ds13}.  The 
effects of a 55,500 \Ms\ thermonuclear SN on a protogalaxy and its surrounding 
cosmological flows has just been studied by \citet{jet13a}.  For explosions in 
low ambient densities, like those of an \HII\ region formed by the star, they 
found that the SN blows most of the baryons from the protogalaxy to radii of 
$\gtrsim$ 10 kpc, or about ten times the virial radius of the halo.  Some of this 
gas, which is heavily enriched by metals from the SN, later falls back into the 
halo on timescales of 100 Myr.

\begin{figure*}
\begin{center}
\begin{tabular}{cc}
\epsfig{file=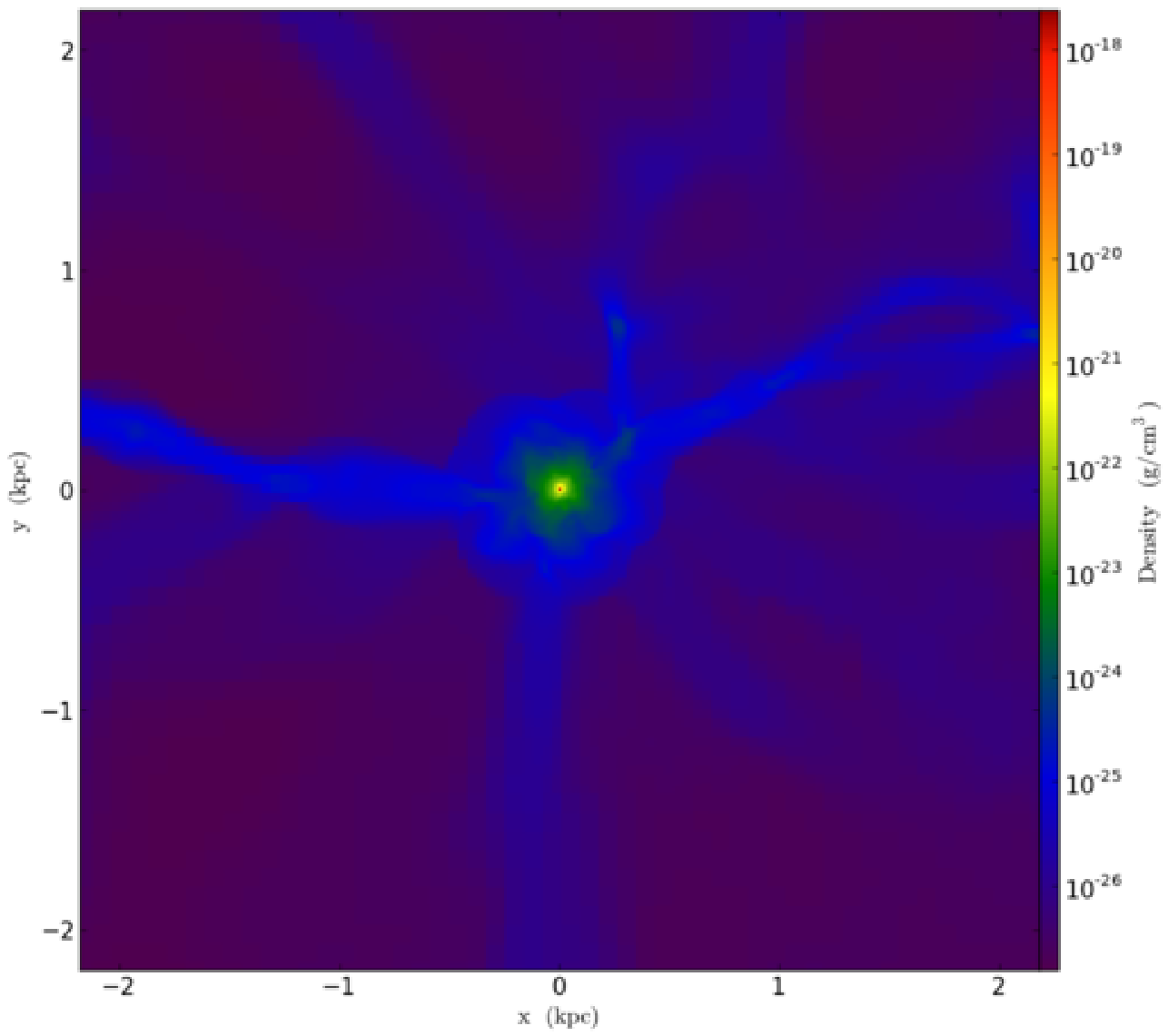,width=0.45\linewidth,clip=} & 
\epsfig{file=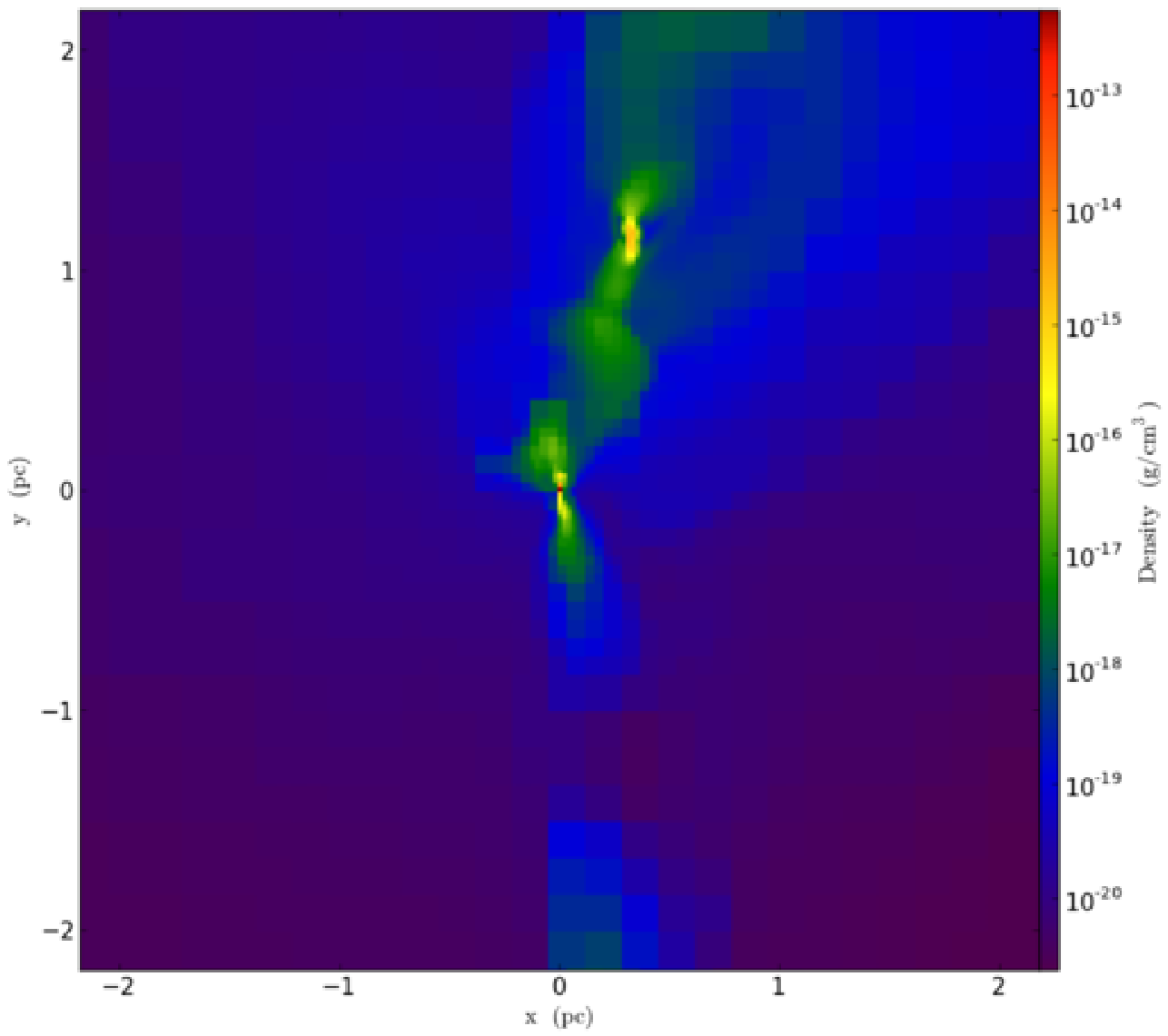,width=0.45\linewidth,clip=} \\
\end{tabular}
\end{center}
\caption{Projections of the 3.2 $\times$ 10$^8$ \Ms\ line-cooled protogalaxy
formed by \textit{Enzo} at $z =$ 15.3.  Left: 1 kpc scale.  Right:  1 pc scale.
Note that a second halo is about to merge with the more massive protogalaxy
on the right.}
\label{fig:enzo1}
\vspace{0.1in}
\end{figure*}

However, past work has shown that many supermassive Pop III stars may not 
form \HII\ regions because of the enormous infall rates at the centers of 
line-cooled halos \citep{jlj12a}.  The star may then have exploded in very high 
ambient densities capable of radiating away most of the energy of the SN as 
bremsstrahlung x-rays in very short times, before the ejecta can be blown from 
the halo \citep{ky05,wet08a} \citep[see also][for studies of less massive SNe in 
early protogalaxies]{vas12}.  This perhaps more likely scenario could not be 
studied by \citet{jet13a} because such losses occur on scales of 0.001 - 10 pc, 
below the resolution limit of their GADGET \citep{gadget1,gadget2} simulation.  
In very dense environments it is not clear if any of the ejecta escapes the halo.  
If not, fallback in the dark matter (DM) potential of the halo could be enormous, 
mixing vigorously with the gas and causing it to cool and fragment into new stars.  
Fallback could also drive super-Eddington accretion in black holes formed by 
other massive fragments in the halo, causing them to grow at much higher rates 
than in the absence of an explosion.

We have now modeled supermassive Pop III SNe in such environments with 
the ZEUS-MP code.  Our simulations bridge all relevant scales of the flow, 
0.001 pc to 10 kpc.  We describe our numerical method in Section 2.  The 
energetics and hydrodynamics of the SNe are examined in Section 3, and we 
calculate the radio signatures of the remnant in Section 4.  The 
implications of our results for the fossil chemical abundance record, starbursts
in early protogalaxies, and SMBH seed growth are discussed in Section 5.

\section{Numerical Method}

We consider explosions in the dense cores of two protogalaxies:  the 4 
$\times$ 10$^7$ \Ms\ atomically-cooled halo from GADGET in \citet{jet13a} 
and a 3.2 $\times$10$^8$ line-cooled halo evolved from cosmological 
initial conditions with the \textit{Enzo} code.  This mass range brackets 
those with which LW protogalaxies are expected to form at $z \sim$ 15. The 
evolution of the SN in spherically-averaged density profiles for these halos is 
then modeled with ZEUS-MP \citep{wn06,wn08a,wn08b}.  As in \citet{wet08a}, 
we evolve the SNe on expanding grids in the DM potential of the protogalaxy.  
Hydrodynamics and nine-species non-equilibrium H and He gas chemistry are 
solved self-consistently to capture the energetics of the SN remnant.  As a 
fiducial case, we examine the dynamics and energetics of the supermassive 
SN in the GADGET halo in detail.

\subsection{Protogalaxy Models / Halo Profiles}

The simulation details for our GADGET halo are described in \citet{jlj11a}; 
here, we discuss the \textit{Enzo} model of our more massive protogalaxy.
\textit{Enzo} \citep{enzo} is an adaptive mesh refinement (AMR) cosmology 
code.  It utilizes an $N-$body adaptive particle-mesh scheme \citep{efs85,
couch91,bn97} to evolve DM and a piecewise-parabolic method for fluid
dynamics \citep{wc84,bryan95}.  A low-viscosity Riemann solver for is used
for capturing shocks, and in these runs we use the recently implemented
HLLC Riemann solver \citep{toro94} for enhanced stability for strong shocks 
and rarefaction waves.  

Like ZEUS-MP, \textit{Enzo} solves nine-species H and He gas chemistry 
and cooling together with hydrodynamics \citep{abet97,anet97}.  To 
approximate the formation of a protogalaxy in a strong LW background, 
we evolve the halo with H$_2$ cooling turned off for simplicity. We initialize 
our simulation volume with gaussian primordial density fluctuations at $z 
=$ 150 with MUSIC \citep{hahn11}, with cosmological parameters from the 
seven-year \textit{Wilkinson Microwave Anisotropy Probe} (\textit{WMAP}) 
$\Lambda$CDM$+$SZ$+$LENS best fit \citep{wmap7}:  $\Omega_{
\mathrm{M}}=$ 0.266, $\Omega_{\Lambda}=$ 0.734, $\Omega_{\mathrm{
b}} = $ 0.0449, $h =$ 0.71, $\sigma_8 = $ 0.81, and $n =$ 0.963.

To capture an atomically-cooled halo in a mass range of 10$^8$ - 10$^9$
\Ms\ by $z \sim$ 15, we use a 2 Mpc simulation box with a resolution of 
1024$^3$.  This yields DM and baryon mass resolutions of 59.7 and 11.6 
$h^{-1}$ \Ms\, respectively.  We use a maximum refinement level $l$ = 13, 
and refine the grid on baryon overdensities of 3 $\times$ 2$^{-0.2l}$.  We 
also refine on a DM overdensity of 3 and resolve the local Jeans length 
with at least four zones at all times to avoid artificial fragmentation during 
collapse \citep{true97}.  If any of these three criteria are met in a given cell, 
the cell is flagged for further refinement.  We show an image of the 
protogalaxy at $z =$ 15.3 in Fig.~\ref{fig:enzo1}.  At this redshift it has 
reached a mass of 3.2 $\times$ 10$^8$ \Ms\ and is about to merge with 
another halo, as shown in the right panel of Fig.~\ref{fig:enzo1}.  

\begin{figure}
\plotone{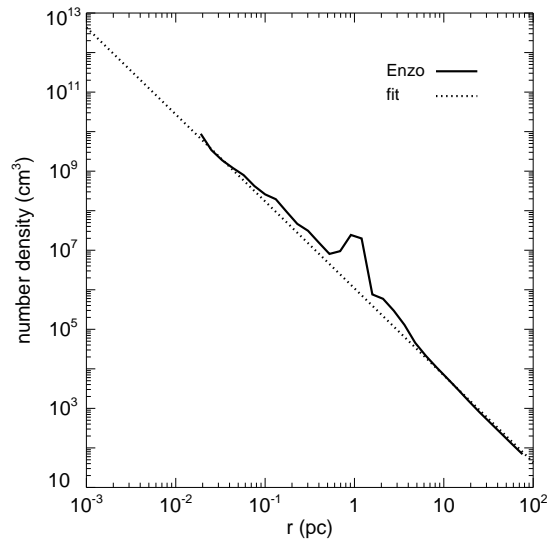} 
\caption{Spherically-averaged baryon density profile for the 3.2 $\times$ 10$
^8$ \Ms\ protogalaxy evolved in \textit{Enzo} together with our fit to this halo
given by Equation \ref{eq:rho2}.  Note that the subsidiary halo in the right 
panel of Fig.~\ref{fig:enzo1}, which is about to merge with the protogalaxy 
and appears in the profile as the bump in density at $\sim$ 1 pc, is excluded 
from our fit.}
\vspace{0.1in}
\label{fig:enzo2}
\end{figure}

The spherically-averaged density profile of the GADGET halo is well
approximated by
\vspace{0.075in}
\begin{equation}
n(r) = 10^3 \left(\frac{r}{10 \; \mathrm{pc}}\right)^{-2} \mathrm{cm}^{-3}.
\vspace{0.075in}
\label{eq:rho}
\end{equation}
The density profile of the more massive protogalaxy evolved in \textit{Enzo}, 
together with our fit for this halo, are shown in Fig.~\ref{fig:enzo2}.  We find 
that it is fit well by 
\vspace{0.075in}
\begin{equation}
n(r) = 6417.1 \left(\frac{r}{10 \; \mathrm{pc}}\right)^{-2.2} \mathrm{cm}^{-3}.
\vspace{0.075in}
\label{eq:rho2}
\end{equation}
The density peak at $\sim$ 1 pc in the \textit{Enzo} profile is due to the 
smaller halo that is about to merge with the protogalaxy in the right panel 
of Fig.~\ref{fig:enzo1}.  It is not included in Eq. \ref{eq:rho2} because it
presents a relatively small solid angle to the oncoming shock and will not
affect its overall dynamics.  We take the gas velocities of both halos to be 
zero for simplicity, their temperatures to be 8500 K (GADGET) and 12,000 
K (\textit{Enzo}), and their mass fractions to be 76\% H and 24\% He.

\subsection{SN Blast Profiles}

\begin{figure*}
\begin{center}
\begin{tabular}{cc}
\epsfig{file=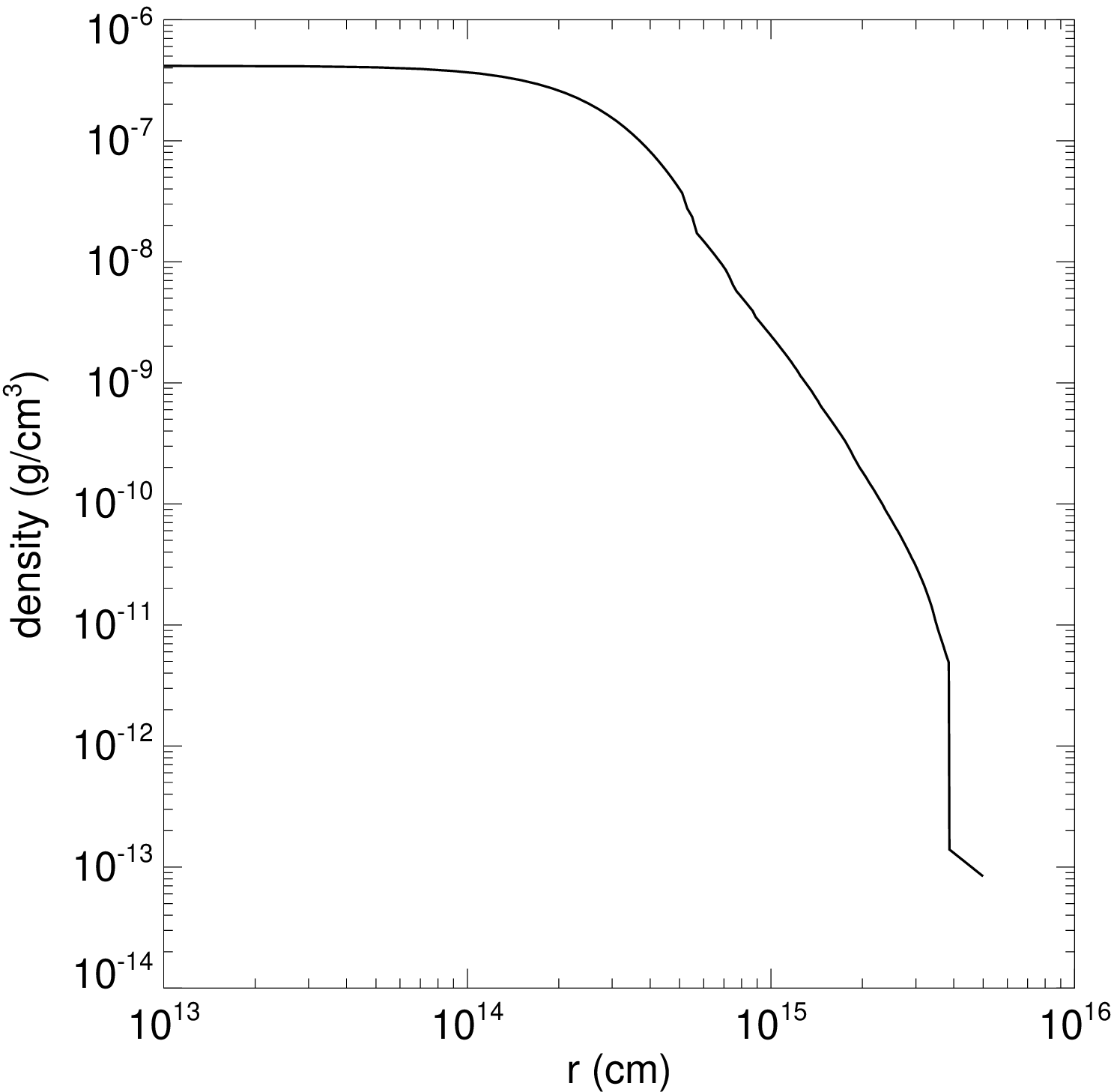,width=0.45\linewidth,clip=} & 
\epsfig{file=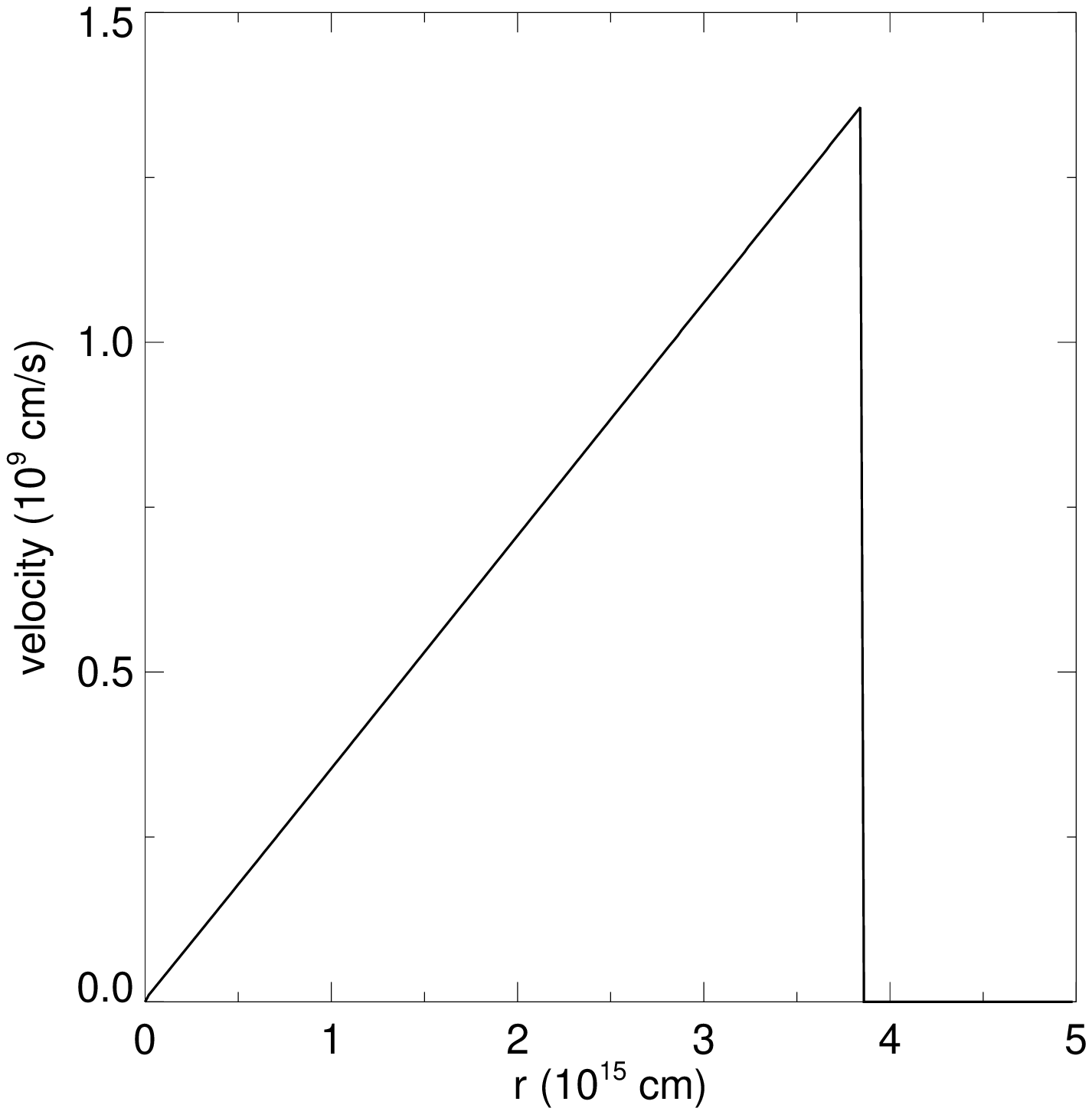,width=0.45\linewidth,clip=} \\
\end{tabular}
\end{center}
\caption{Profiles for the supermassive Pop III SN and the surrounding halo as 
initialized in ZEUS-MP.  Left:  densities; right:  velocities.  The $r^{-2}$ density
envelope of the protogalaxy (for the case of no \HII\ region) is visible beyond 
the shock.  At this stage, the SN is a free expansion.}
\label{fig:initprof}
\end{figure*}

Our initial blast profile is the 10$^{55}$ erg thermonuclear SN of a 55,500 
\Ms\ star from \citet{wet12e} \citep[see also][]{heg13}.  The explosion was
evolved in the \textit{Kepler} and RAGE codes \citep{Weaver1978,
Woosley2002,rage,fet12} out to 2.9 $\times$ 10$^6$ s, after breakout from 
the star but well before it has swept up its own mass. The SN ejects 23,000 
\Ms\ of metals \citep[and perhaps molecules and dust;][]{cl08,cd09,cd10,
dc11,gall11} into the IGM.  At this stage it is a free expansion, and its 
potential energy is a small fraction of its total energy so it is not necessary 
to account for its self gravity.  We show profiles for the explosion in 
Fig.~\ref{fig:initprof}.  Densities, velocities and internal energies from RAGE 
are mapped onto a uniform one-dimensional (1D) spherical polar coordinate 
grid in ZEUS-MP using a simple linear interpolation of the logarithm of the 
quantity versus log radius. The radius of the shock is 3.9 $\times$ 10$^{15}
$ cm and the inner and outer grid boundaries are zero and 5.0 $\times$ 
10$^{15}$ cm, respectively.  The coordinate mesh has 250 zones.  

\subsection{Nonequilibrium H/He Gas Chemistry}

We evolve H, H$^+$, He, He$^+$, He$^{++}$, H$^-$, H$^{+}_{2}$, H$_{2}$, 
and e$^-$ with the thirty-reaction rate network described in \citet{anet97} in
order to tally energy losses in the shock as it sweeps up and heats baryons.  
Cooling by collisional excitation and ionization of H and He, recombinations, 
inverse Compton (IC) scattering from the cosmic microwave background 
(CMB), and free-free emission by bremsstrahlung x-rays are included.  The 
rates at which these processes occur depend on the chemical species in the 
shocked gas, which in turn are governed by the reaction network.  We 
exclude H$_2$ cooling because high temperatures destroy these fragile 
molecules in the shock.  At early stages of the explosion we also deactivate 
H and He line cooling because x-rays from the shock ionize these species 
in the swept-up gas.  These cooling channels are switched on when the 
shock temperature falls to 10$^6$ K.  

\subsection{Static DM Potential}

Rather than spherically-averaging the DM distribution of the protogalaxy in 
GADGET and mapping its gravitational potential into ZEUS-MP, we simply
take its potential to be that required to keep the density profile in Equation 
(\ref{eq:rho}) in hydrostatic equilibrium on the grid.  The mass associated 
with this potential is nearly identical to that of the halo.  We interpolate this 
precomputed potential onto the grid at the beginning of the run and onto 
each new grid thereafter, but the potential itself never evolves.  Assuming 
the potential to be static is an approximation because unlike the models of 
\citet{wet08a}, dynamical times for supermassive SNe in protogalaxies are 
comparable to merger and accretion timescales in the protogalaxy.  It has 
also been shown that Pop III SNe in halos with DM cusps can alter the 
structure of the cusp \citep{ds11a}, and hence the potentials in which they
themselves evolve.  

\subsection{Expanding Grid}

\begin{figure*}
\begin{center}
\begin{tabular}{cc}
\epsfig{file=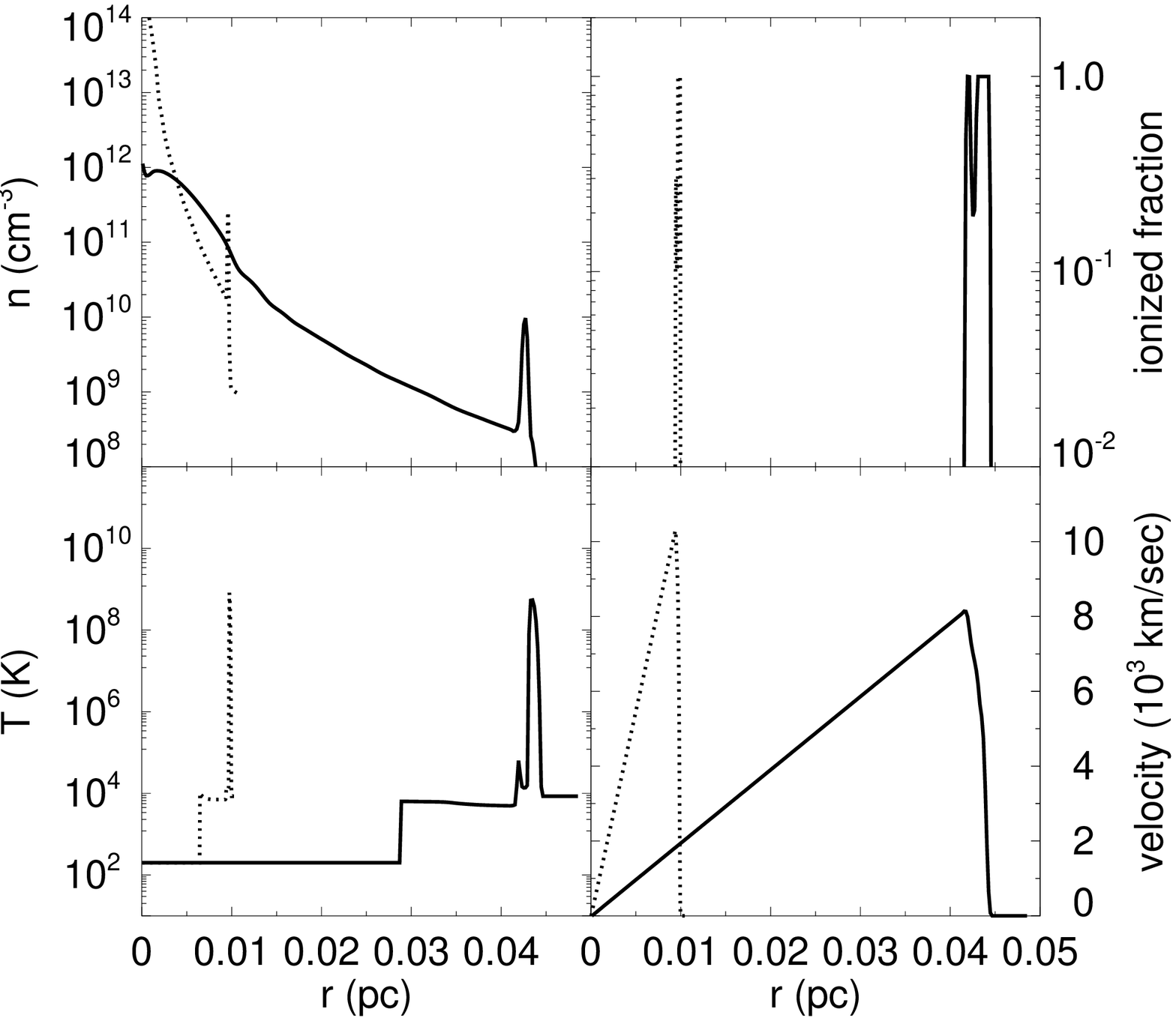,width=0.45\linewidth,clip=} & 
\epsfig{file=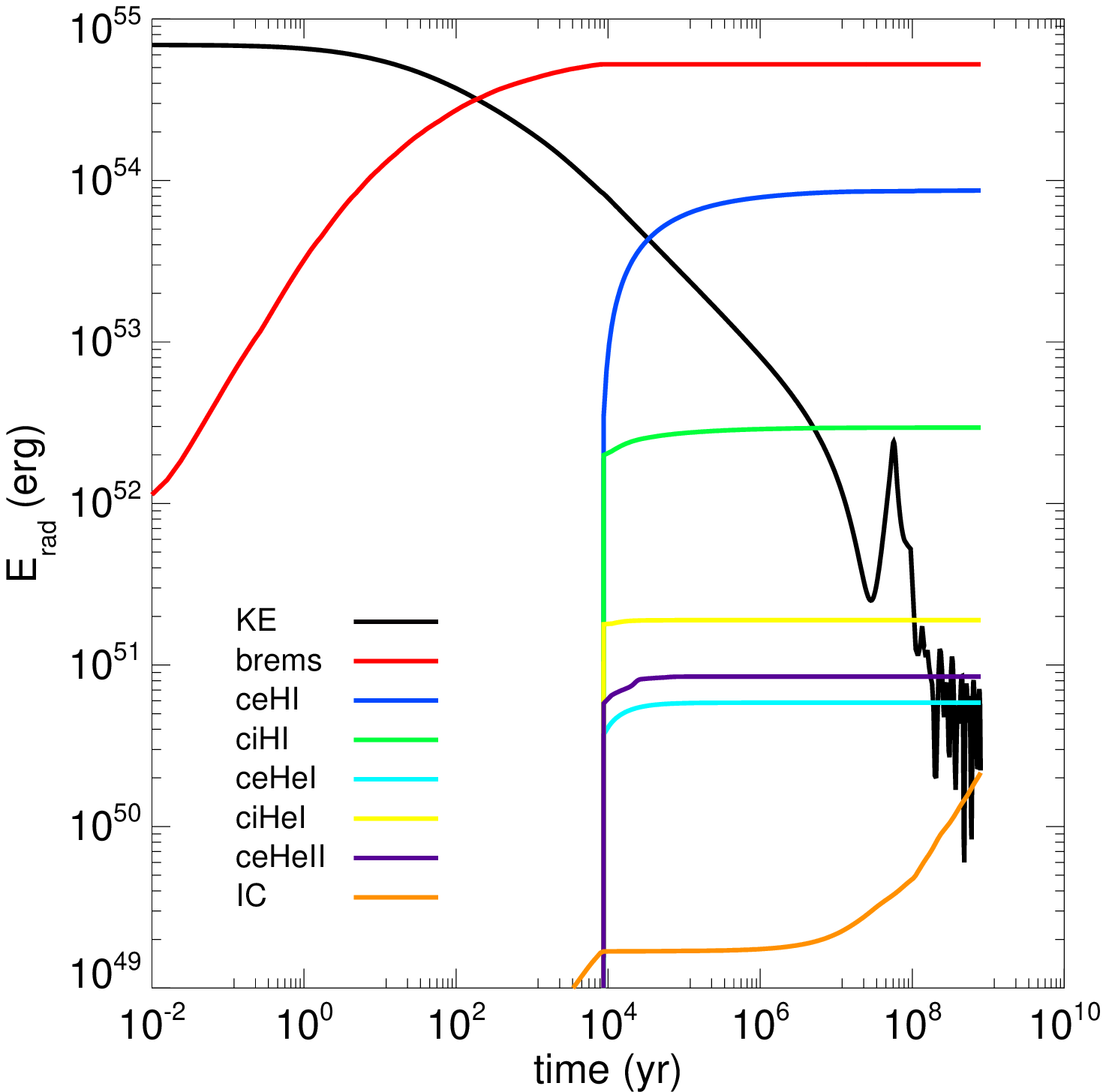,width=0.45\linewidth,clip=} \\
\end{tabular}
\end{center}
\caption{Left:  The interaction of the free expansion with the dense protogalactic 
envelope. Dotted line: 0.658 yr.  Solid line:  4.04 yr.  Right:  cumulative radiative 
losses from the SN remnant in the 4 $\times$ 10$^7$ \Ms\ protogalaxy versus 
time.  KE is the kinetic energy of the remnant and the other plots are losses due 
to bremsstrahlung x-rays (brems), collisional excitation of H (ceHI), collisional 
ionization of H (ciHI), collisional excitation of He (ceHeI), collisional ionization of 
He (ciHeI), collisional excitation of He$^+$ (ceHeII) and cooling due to 
upscattering of CMB photons (IC).  Cooling due to collisional ionization of He$^+$ 
is negligible and not shown.}
\label{fig:exp}
\end{figure*}

\begin{figure*}
\begin{center}
\begin{tabular}{cc}
\epsfig{file=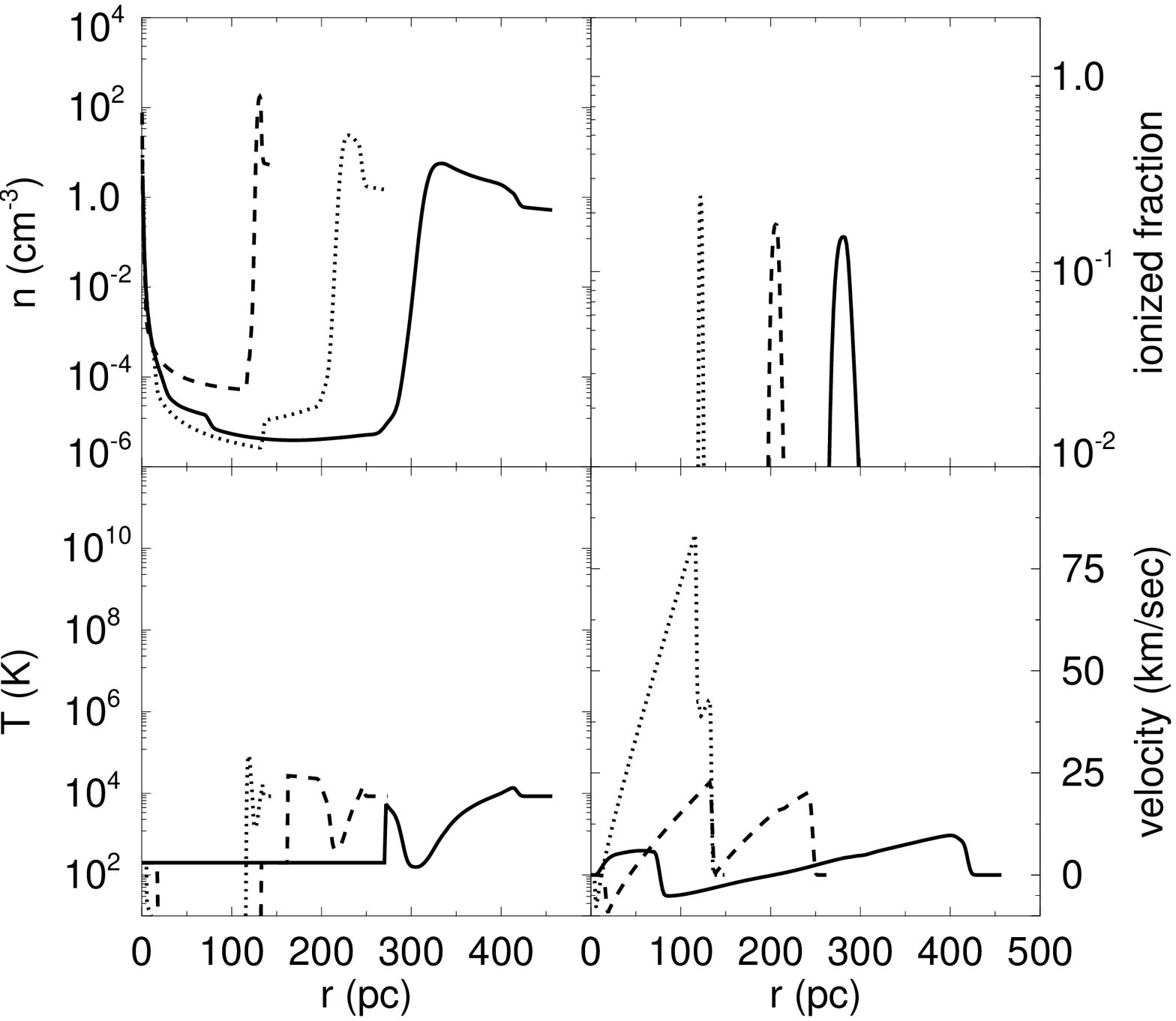,width=0.45\linewidth,clip=} & 
\epsfig{file=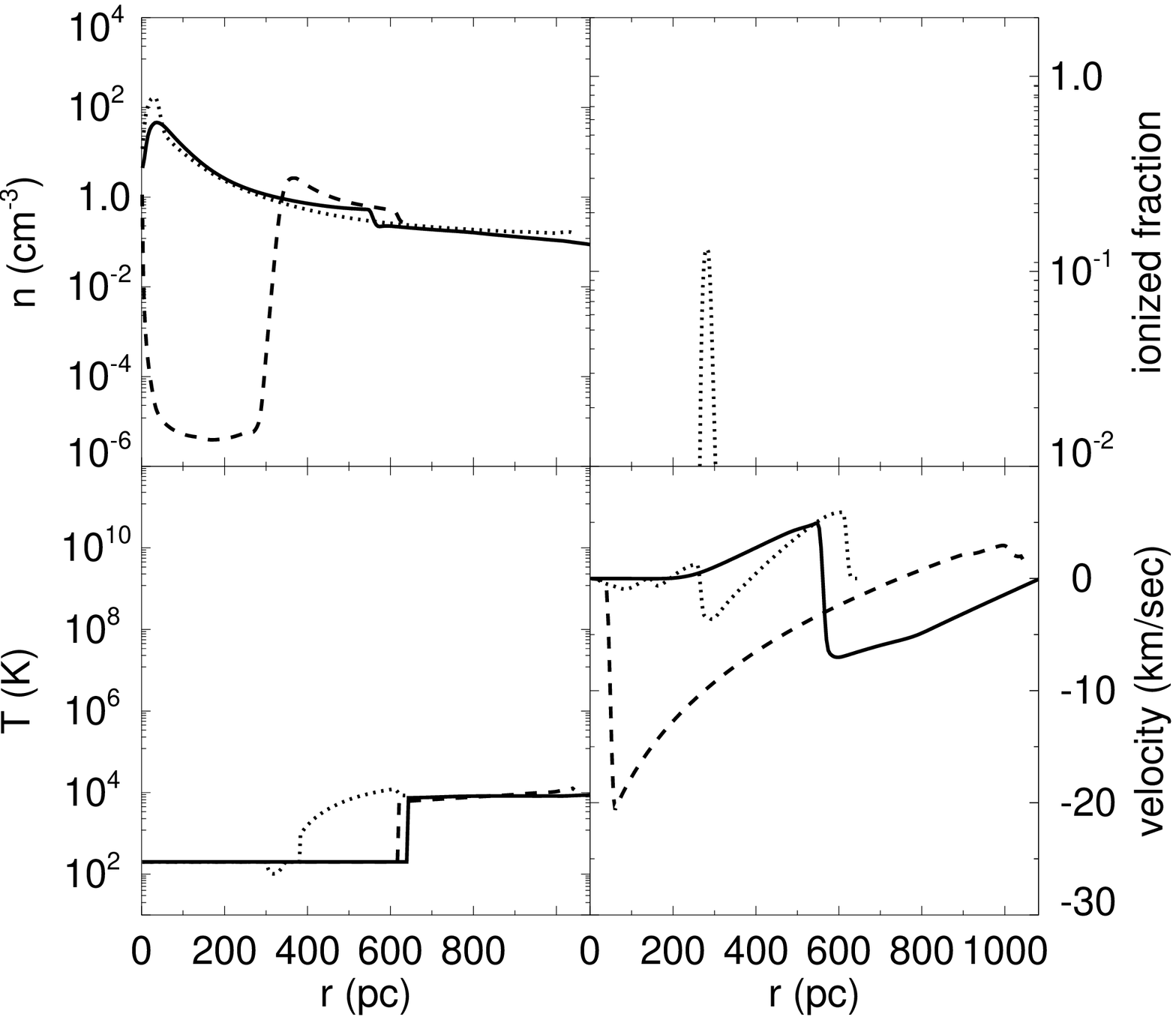,width=0.45\linewidth,clip=} \\
\end{tabular}
\end{center}
\caption{Left:  Detachment of the reverse shock from the forward shock and its 
propagation into the interior of the remnant at 1.19 Myr (dashed line), 4.68 Myr 
(dotted line), and 13.5 Myr (solid line).  Right:  fallback into the protogalaxy.  
Dashed line:  26.2 Myr; dotted line:  58.7 Myr; solid line:  95.2 Myr.}
\vspace{0.1in}
\label{fig:revshk}
\end{figure*}

When the shock reaches a predetermined distance from the outer edge of 
the grid, a new grid is created and densities, velocities, energies and species 
mass fractions for the flow are mapped onto it.  The ambient density of the 
halo, which is precomputed in a table that is binned by radius, is imposed on 
the expanding flow as time-dependent updates to the outer boundary 
conditions.  The density that is assigned to the outer boundary is interpolated 
from the tabled values bracketing the radial bin into which the grid boundary 
falls at the given time step.  This density then migrates inward toward the 
shock over time as it is mapped onto the grid at greater and greater distances 
from the outer boundary in subsequent regrids.  In this manner the shock 
always encounters the proper halo densities as it grows by many orders of 
magnitude in radius.  At the beginning of a run we allocate the first 80\% of 
the grid to the SN profile and the remainder to the surrounding halo.  

A new grid is calculated when the shock crosses into the outer 10\% of the 
current mesh.  At a given time step, the maximum gas velocity in this region 
and its position are determined.  A homologous velocity $v(r) = \alpha r$ is 
then assigned to each grid point such that the velocity of the grid at the 
radius of maximum gas velocity in the outer 10\% of the mesh is three times 
this maximum.  The factor of 3 guarantees that the flow never reaches the 
outer edge of the grid, where boundary conditions are reset every time step 
to ensure that the SN always encounters the correct protogalactic density, 
as described above.  A new radius is then computed for each grid point from 
its present radius, the velocity it is assigned from the homologous profile, and 
the current time step.  This procedure preserves the total number of grid 
points and ensures that they are uniformly spaced between the new inner 
and outer boundaries.

\section{Evolution of the SN}

\subsection{Free Expansion}

As the shock initially plows into the dense envelope it heats it to nearly 10$
^9$ K, as we show in the left panel of Fig.~\ref{fig:exp}. At these temperatures 
x-ray emission dominates energy losses from the shock and the ejecta rapidly 
loses energy, as shown in the right panel of Fig.~\ref{fig:exp}.  By $\sim$ 10$
^4$ yr bremsstrahlung losses have removed over 90\% of the kinetic energy 
of the remnant.  By this time the shock has cooled to $\sim$ 10$^6$ K, and 
collisional excitation and ionization losses in H and He are activated in the gas.  
The shock cools to $\sim$ 10$^4$ K in less than a Myr thereafter, because 
energy losses taper off by that time in Fig.~\ref{fig:exp}.  Losses by excitation 
and ionization of H and He in the shocked, swept-up gas eventually amount to 
$\sim$ 1\% of the original kinetic energy of the SN, with collisional excitation of 
H being the dominant cooling channel.  During this time the remnant radiates 
large Ly-$\alpha$ fluxes.

\subsection{Formation of the Reverse Shock / Onset of Mixing}

As seen in the double peak in both ionization fractions and temperatures in 
Fig.~\ref{fig:exp}, a reverse shock forms in the free expansion at early times
but is not strong enough to fully detach from the forward shock or completely 
ionize the gas.  It becomes stronger as the remnant sweeps up more of the 
halo, eventually heating to over 10$^6$ K and emitting large x-ray fluxes. 
The SN plows up approximately its own mass for each pc it expands in the 
halo, and when it reaches a radius of $\sim$ 150 pc the reverse shock fully 
detaches from the forward shock and backsteps into the ejecta, as shown at 
1.19 and 4.68 Myr in the left panel of Fig.~\ref{fig:revshk}.  The forward shock 
decelerates from $\sim$ 80 km s$^{-1}$ to 13 km s$^{-1}$. 

The arrival of the reverse shock at the center of the halo and its subsequent 
rebound is visible in the velocity peak at $\sim$ 75 pc at 13.5 Myr.  By this 
stage the bulk of the remnant lies in a dense shell from 300 - 400 pc and it 
has evacuated the interior to very low densities.  Essentially all radiative 
cooling has halted at this point because gas temperatures have fallen below 
10$^4$ K everywhere, so the forward and reverse shocks now propagate 
adiabatically.  We note that the retreat of the reverse shock through the 
dense ejecta shell would likely trigger Rayleigh-Taylor instabilities in three
dimensions and mix metals with gas from the halo.  Turbulent mixing driven
by fallback and inflow from filaments would then further distribute metals 
throughout the interior of the protogalaxy at later times.

\subsection{Fallback}

\begin{figure}
\plotone{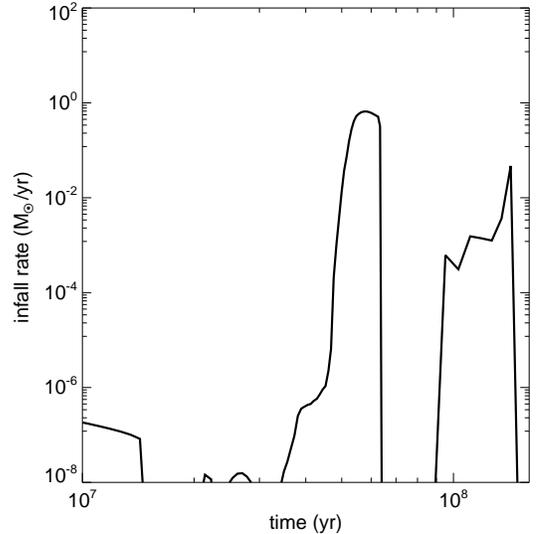} 
\caption{Accretion rates at the center of the 4 $\times$ 10$^7$ \Ms\ protogalaxy 
due to fallback.  The first episode of fallback, from 5 - 6 $\times$ 10$^7$ yr, has 
the largest mass flux, $\sim$ 1 \Ms\ yr$^{-1}$.  Later bouts of accretion are due 
to subsequent cycles of adiabatic expansion and collapse of baryons in the halo.} 
\vspace{0.1in}
\label{fig:fallback}
\end{figure}

By 26.2 Myr, when the remnant has grown to $\sim$ 600 pc, the gravitational 
potential of the halo has halted its expansion, and all but its outermost layers 
fall back toward the center of the protogalaxy as shown in the right panel of 
Fig.~\ref{fig:revshk}.  The reverse shock rebounding from the center of the 
protogalaxy collides with the inner surface of the collapsing shell and is again 
driven back into the center of the halo.  This shock reverberates between the 
shell and the center of the protogalaxy with increasing frequency as the ejecta 
crushes the low-density cavity into the center of the halo.  The shell reaches 
the center of the halo by 58.7 Myr, raising gas densities to $\sim$ 200 cm$^{
-3}$ there.  As shown in Fig.~\ref{fig:fallback}, infall rates through the central 
50 pc of the halo, as tallied by \vspace{0.05in}
\begin{equation}
\dot{m} \; = \; 4 \pi r^2 \rho v_{\mathrm{gas}}, \vspace{0.05in}
\end{equation}
reach $\sim$ 1 \Ms\ yr$^{-1}$ and persist for nearly 10$^7$ yr, so most of the 
baryons originally interior to the virial radius of the protogalaxy eventually fall 
to its center.  Meanwhile, the momentum of the uppermost layers of the 
remnant lift the outer layers of the protogalaxy to altitudes of 1 - 1.5 kpc but 
they soon settle back down onto the halo.

After this first episode of fallback, the ejecta and baryons rebound adiabatically
back out into the protogalaxy, as shown in Fig.~\ref{fig:revshk} at 95.2 Myr. 
The leading edge of the outward flow is visible as the ripple in density and the 
velocity peak at $\sim$ 600 pc. As the ejecta propagates outward, it again stalls 
in the DM potential of the halo and falls back toward the center.  In our models 
this cycle continues for several hundred Myr and is manifest as the oscillations 
in KE at late times in the right panel of Fig.~\ref{fig:exp}. In reality, accretion from 
filaments and mergers with other halos would disrupt this motion as they agitate 
gas at the center of the protogalaxy.  Furthermore, departures from spherical 
symmetry in real protogalaxies, or off-center explosions, could allow more ejecta 
to flow out into the IGM, especially if UV flux from the SN progenitor opens 
channels of low column density out of the halo through which metals can escape.
If metals and dust mix completely with the baryons in the halo, as our results 
suggest, the metallicity of the protogalaxy will abruptly rise to 0.1 \Zs\ after just 
one explosion in 50 - 100 Myr.  

The fact that the protogalaxy traps most metals from explosions this energetic 
is counter to what might naively be expected from simple binding energy 
arguments.  The binding energy $E_{\mathrm{B}}$ of the gas in the halo can 
be approximated by that of an isothermal sphere, \vspace{0.05in}
\begin{equation}
E_{\mathrm{B}} = \int^{R_{\mathrm{vir}}}_0 \frac{GM_{\mathrm{encl}}}{r} 4\pi r^2 \rho_\mathrm{B}(r) dr,
\vspace{0.05in}
\end{equation}
where 
\begin{equation}
M_{\mathrm{encl}} = \int^r_0 4\pi {r'}^2 \left(\rho_{\mathrm{DM}}(r') + \rho_\mathrm{B}(r')\right) dr'
\vspace{0.05in}
\end{equation}
and 
\begin{equation}
\rho_{\mathrm{DM}} = \frac{\Omega_\mathrm{M} - \Omega_\mathrm{B}}{\Omega_\mathrm{B}} \rho_\mathrm{B}.
\vspace{0.05in}
\end{equation}
If we take $\Omega_\mathrm{M} =$ 0.266 and $\Omega_{\mathrm{B}} =$ 
0.0449, set the virial radius $R_{\mathrm{vir}}$ of the halo to be 1 kpc, and
use Equation \ref{eq:rho} for $\rho_{\mathrm{B}}$, we obtain $E_{\mathrm{
B}} \sim$ 8.5 $\times$ 10$^{53}$ erg, which far less than the energy of the 
SN.  But the gas is not blown out of the protogalaxy because the ejecta loses 
$>$ 95\% of its kinetic energy to bremsstrahlung x-rays by the time it reaches 
1\% of the virial radius of the halo.  In fact, after radiative losses the kinetic
energy of the remnant is very close to the binding energy of the gas to the 
dark matter, and this is why it only expands to $R_{\mathrm{vir}}$ before 
falling back into the halo.  The DM potential of the protogalaxy slows down 
the growth of the remnant, never allowing it to become a momentum
conserving snowplow.  The remnant retains far more of its energy in \HII\ 
regions, and blows all the gas from the protogalaxy as shown by \citet{jet13}.

\begin{figure*}
\begin{center}
\begin{tabular}{cc}
\epsfig{file=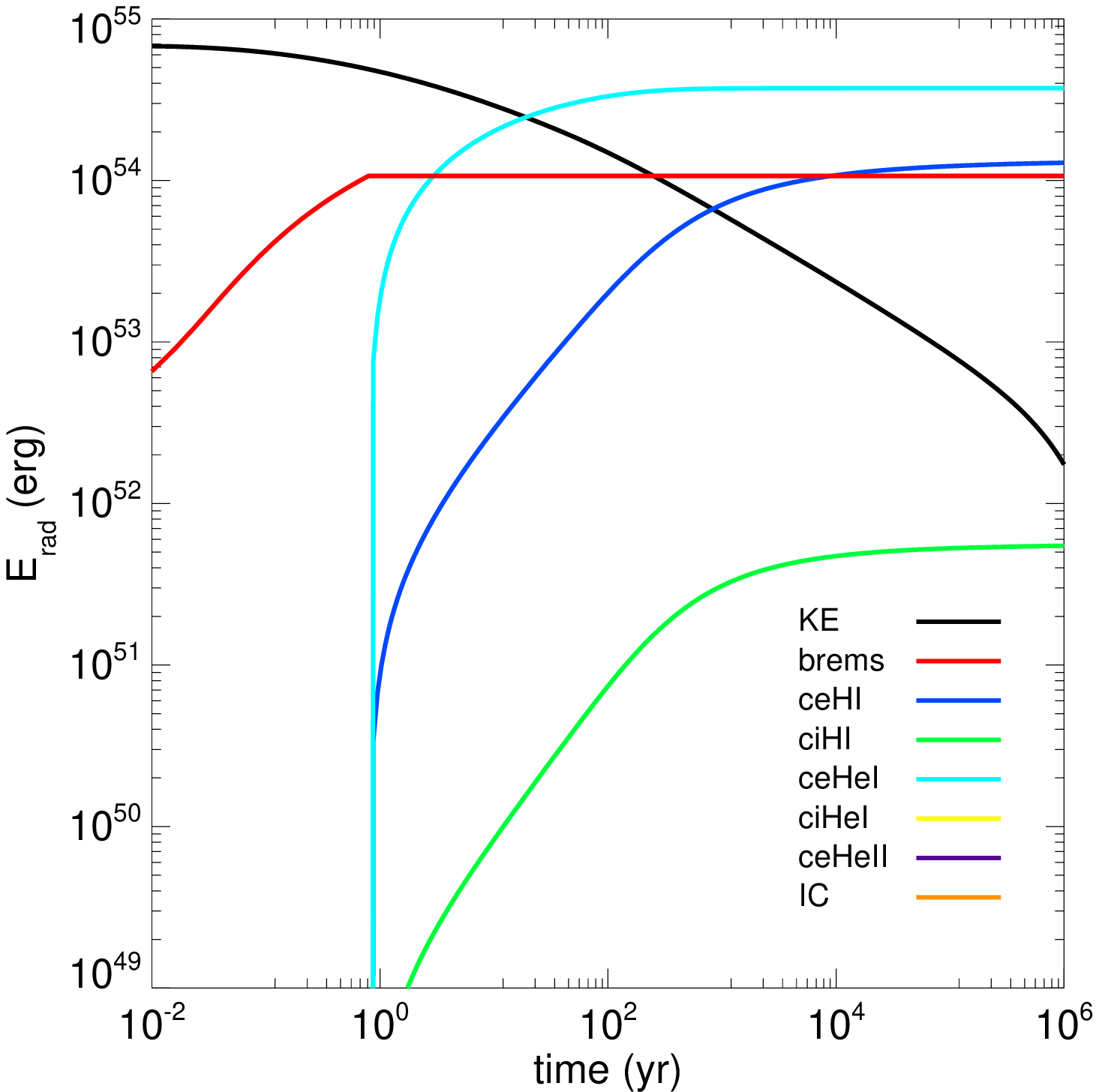, width=0.45\linewidth,clip=}
\epsfig{file=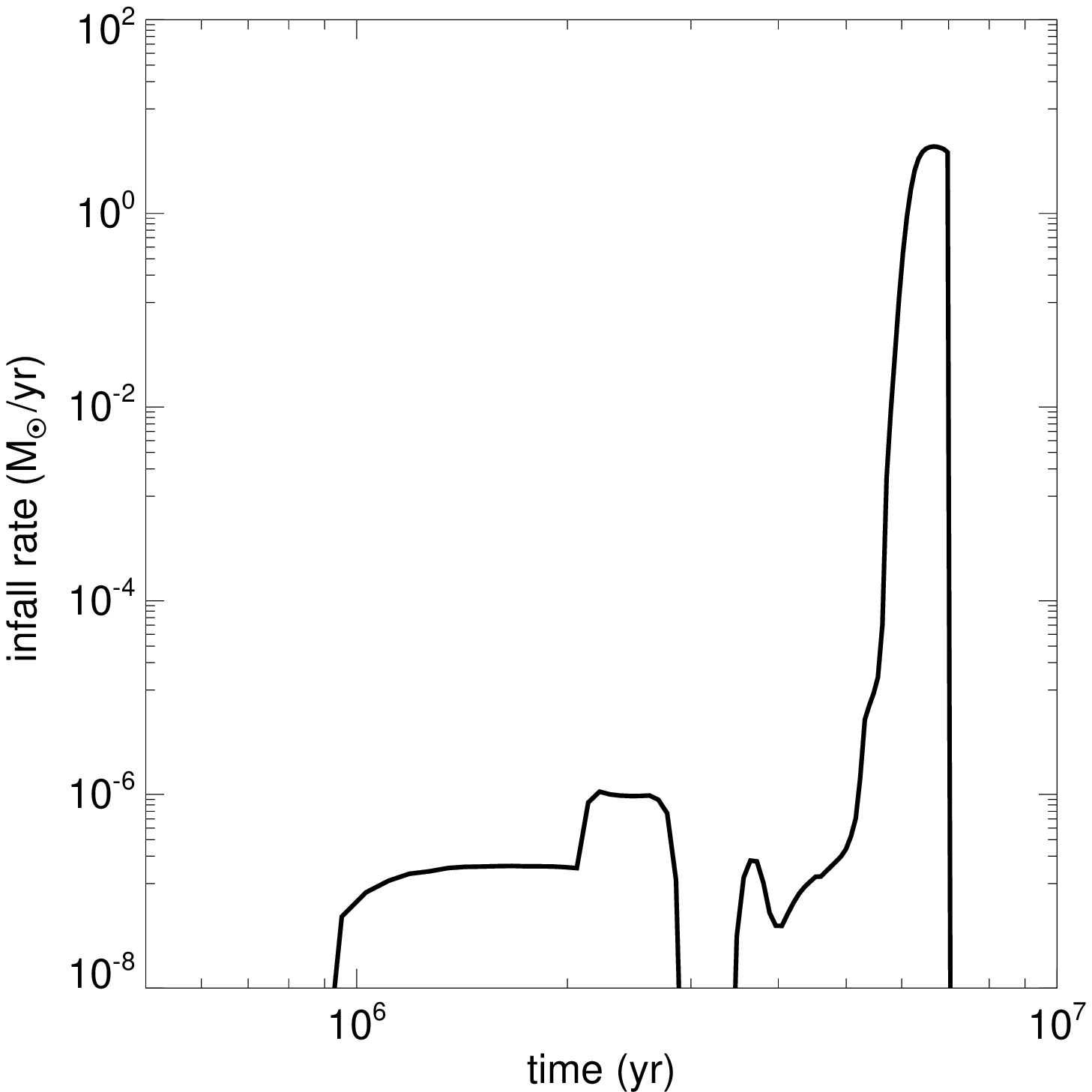, width=0.45\linewidth,clip=}
\end{tabular}
\end{center}
\caption{Left:  cumulative radiative losses from the SN remnant in the 3.2 $\times$ 
10$^8$ \Ms\ protogalaxy versus time.  KE is the kinetic energy of the remnant and 
the other plots are losses due to bremsstrahlung x-rays (brems), collisional 
excitation of H (ceHI), collisional ionization of H (ciHI), collisional excitation of He 
(ceHeI), collisional ionization of He (ciHeI), collisional excitation of He$^+$ (ceHeII) 
and cooling due to upscattering of CMB photons (IC).  Cooling due to ciHeII, ceHeII,
and IC are negligible and not shown.  Right:  infall rates at the center of the 3.2 
$\times$ 10$^8$ \Ms\ protogalaxy due to fallback.}
\label{fig:enzo3}
\vspace{0.1in}
\end{figure*}

We neglect cosmic ray emission from the remnant because the strengths of 
magnetic fields in $z \sim$ 15 protogalaxies, or even the progenitor stars 
themselves, are not well constrained \citep[although see][]{schob12,latif13b}. 
But such emission would simply slow down the expansion of the SN, confining 
it to even smaller radii than those here.  Likewise, non-equilibrium radiative 
cooling by metals in the ejecta, which also transports some internal energy 
out of the remnant, is not included in our simulations.  Consequently, the final 
radii of the explosions in our models should be taken to be upper limits. 

\subsection{Supermassive Pop III SNe and the CMB}

Inverse Compton losses from the remnant are modest, just 1\% of those from 
140 - 260 \Ms\ pair-instability (PI) SNe in $z \sim$ 25 minihalos \citep{wet08a}.  
In those explosions the shock remains hotter at large radii because they occur
in \HII\ regions, so CMB photons are upscattered with greater efficiency in their 
interiors.  In the SNe in this study, IC losses are much lower because the 
remnant cools before enclosing large volumes of CMB photons. The losses are 
small at first but later grow as the shock reaches larger radii.

It is clear that supermassive explosions in \HII\ regions in early protogalaxies 
would leave a larger imprint on the CMB via the Sunyayev-Zeldovich effect than 
the SNe in our models.  Low explosion rates probably prevented such SNe from 
collectively imposing excess power on the CMB on small scales. However, unlike 
PI SNe at earlier epochs, these SN remnants may become large enough to be 
directly resolved by current instruments like the \textit{Atacama Cosmology 
Telescope} (\textit{ACT}) and the \textit{South Pole Telescope} (\textit{SPT}) in 
addition to being detected in the near infrared \citep[NIR;][]{wet12d} by future 
missions such as the \textit{Wide-Field Infrared Survey Telescope} 
(\textit{WFIRST}) and the \textit{Wide-Field Imaging Surveyor for High Redshift} 
(\textit{WISH}).  Calculations of the imprint of these and other types of Pop III 
SNe on the CMB are now underway.

\subsection{SN in More Massive Halos}

Explosions in \ion{H}{2} regions in protogalaxies, as in \citet{jet13}, can eventually
range 10 kpc or more from their host halos, perhaps even contaminating other
nearby protogalaxies with metals.  In contrast, SNe in undisturbed halos briefly 
engulf the protogalaxy but then collapse back into it, with few metals lost to the 
IGM.  Our fiducial halo is on the low end of the mass scale for atomically cooled 
protogalaxies so explosions in more massive structures will be confined to even
smaller radii, as we find for the SN in our 3.2 $\times$ 10$^8$ \Ms\ halo.  This 
explosion evolves in basically the same manner as in the less massive halo but  
with two important differences.

First, because the shock initially plows into much higher densities it cools more 
quickly, radiating fewer x-rays and allowing H and He line cooling to activate at 
much earlier times.  As we show in the right panel of Fig.~\ref{fig:enzo3}, He line
losses dominate cooling in the remnant rather than x-rays. Second, because the
shock cools more rapidly the SN ejecta only expands to a radius of 80 pc before
falling back into the halo.  Consequently, this explosion will be less efficient at 
enriching gas in the protogalaxy although turbulent mixing by cosmological flows 
along filaments will still enhance this process.  As shown in the right panel of 
Fig.~\ref{fig:enzo3}, peak fallback rates in this protogalaxy are nearly ten times 
those in the less massive halo but last for only 2 Myr instead of 10 Myr.  About 
the same amount of gas collapses back to the center of the halo in this first 
episode of fallback, fueling the rapid growth of any SMBH seeds there.

\begin{figure*}
\begin{center}
\begin{tabular}{cc}
\epsfig{file=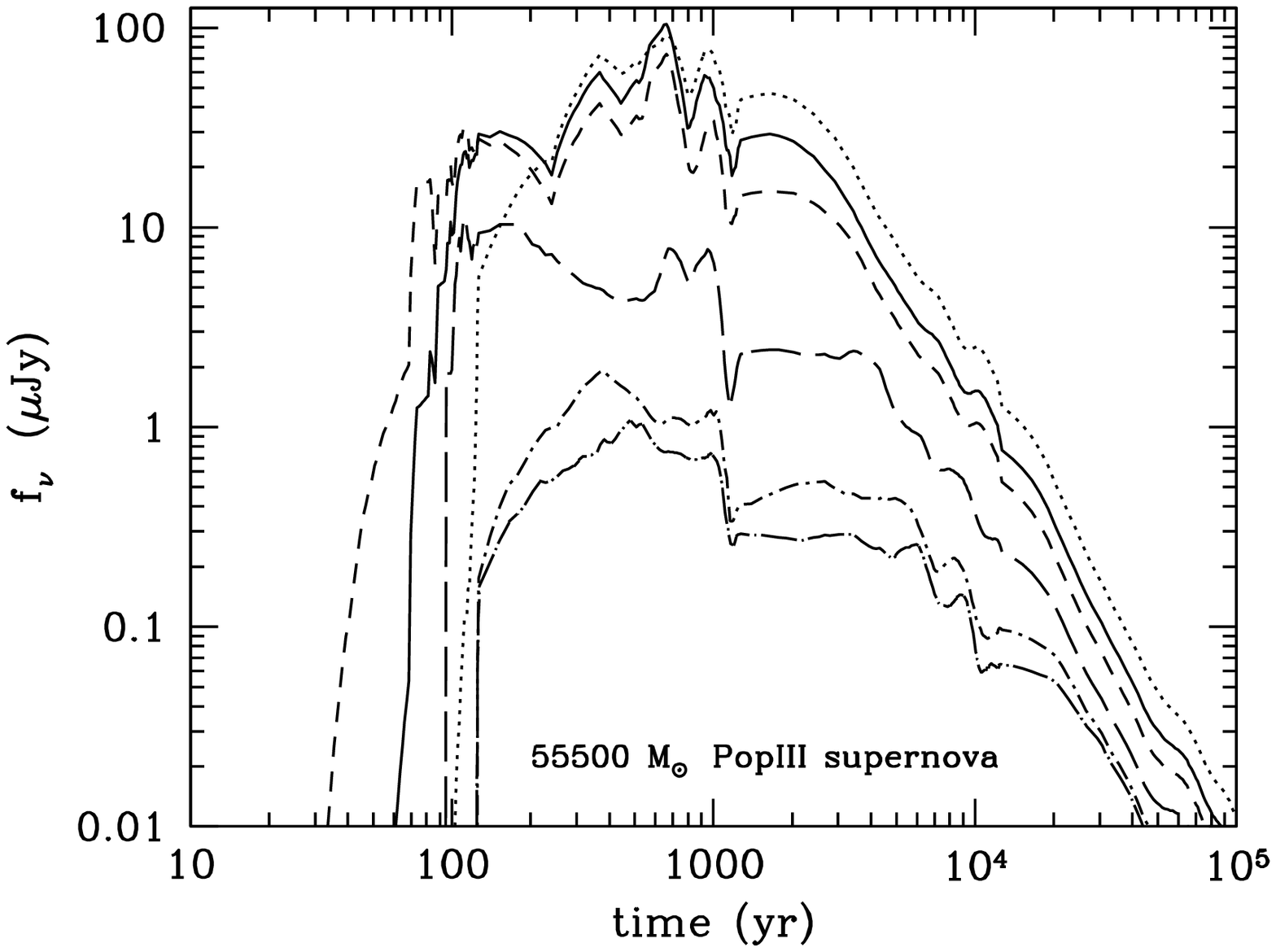, width=0.45\linewidth,clip=}
\epsfig{file=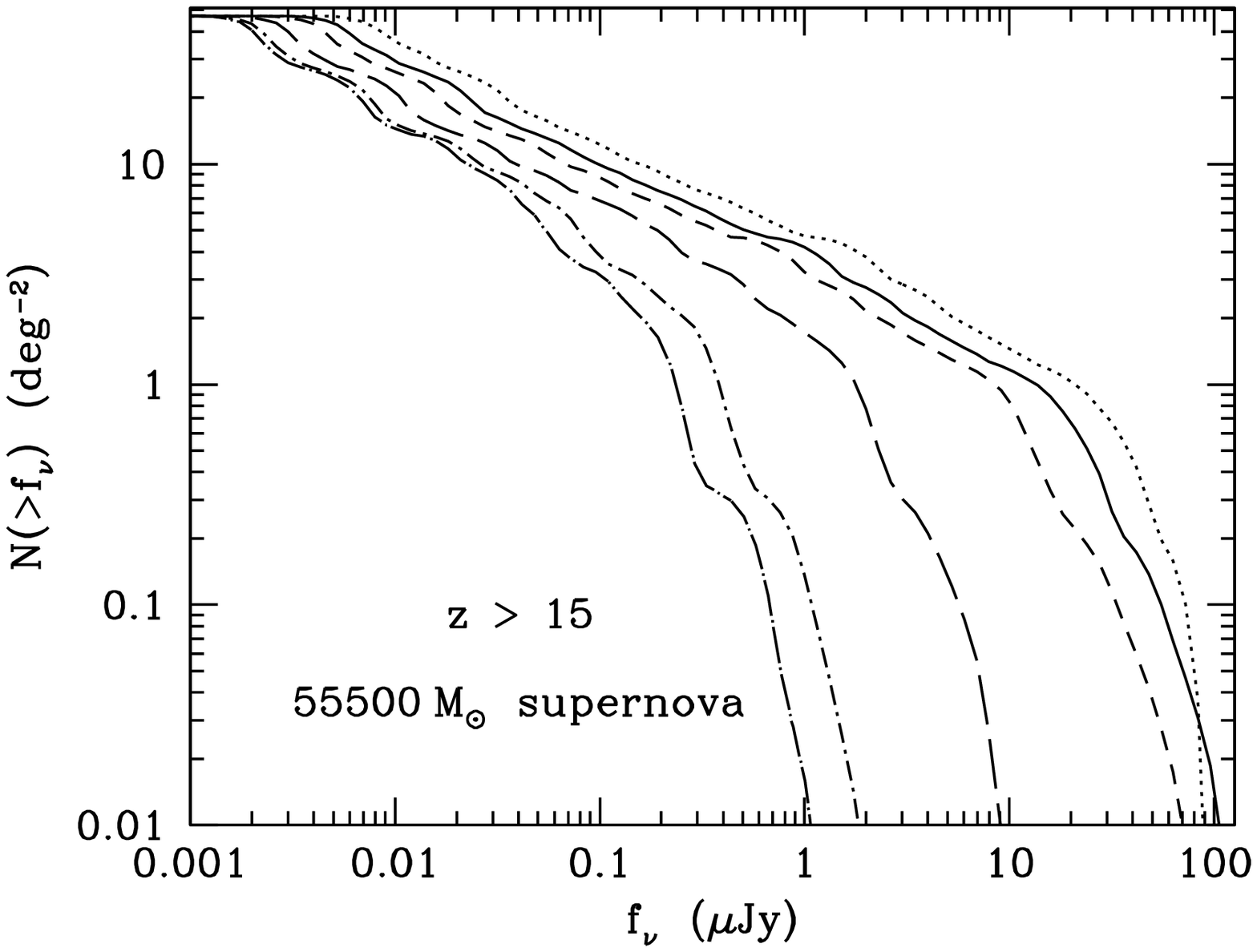, width=0.45\linewidth,clip=}
\end{tabular}
\end{center}
\caption{Left:  radio light curves for a remnant in a $4\times10^7$ \Ms\ halo at $z
=15$ (observer's frame).  The curves correspond to the bands:\ 0.5 (dotted), 1.4 
(solid), 3 (short-dashed), 10 (long-dashed), 25 (dot short-dashed) and 35 (dot 
long-dashed) GHz.  Right: the corresponding number counts above a given 
observed flux, assuming one SN per halo (the counts must be scaled down 
proportionately for lower supermassive SN rates).}
\label{fig:Mh4e7_lightcurves}
\end{figure*}

\section{Radio Signatures of Supermassive Pop III SNe}

Shocks in SN remnants accelerate electrons to relativistic velocities, giving rise to 
radio synchrotron emission. We compute the emission spectrum of the SN in both
protogalaxies, including both free-free and synchrotron radiation, as in \citet
{mw12}. Specifically, the electron distribution is taken to be a power law in energy 
$E=(\gamma-1)m_ec^2$, $n(E) \propto E^{-p_e}$, with an average spectral index 
$p_e=2$.  Energy equipartition between the relativistic electrons and the magnetic 
field is assumed, allowing for a fraction $f_e=0.01$ of the thermal energy to go into 
the relativistic electrons.  The range in the relativistic $\gamma$ factor is limited by 
post-shock energy losses, primarily plasmon excitation at the low energies and 
synchrotron cooling at high energies.  Free-free absorption and synchrotron 
self-absorption are included.

The evolution of the spectrum of the SN in the lower-mass GADGET halo in the 
observer's frame for a source at $z = 15$ is shown in the left panel of 
Fig.~\ref{fig:Mh4e7_lightcurves} for several radio bands, including those accessible 
to the Square Kilometer Array\footnote{\texttt{www.skatelescope.org}} (SKA) and 
ASKAP\footnote{\texttt{www.atnf.csiro.au/projects/askap}}.  The peak radio fluxes 
are within the sensitivity range of the eVLA\footnote{\texttt{www.aoc.nrao.edu/evla}} 
and eMERLIN\footnote{\texttt{www.jb.man.ac.uk/research/rflabs/eMERLIN.html}}.
The delay in the flux is due to the energy losses which pinch off the relativistic 
electron population at early times. Once the emission starts, the flux rises sharply, 
with a doubling time of months in the observer's frame.  The flux is sharply peaked 
around 3~GHz during the rise, followed by radiation emerging at 1.4~GHz then 10
GHz.  The signature is unusual and suggests an observing strategy of periodic
surveys with observations carried out at intervals of several months to a few years, 
searching for such narrowly peaked evolving sources.  At late times, the source is 
red in the radio with most of the flux at the lower frequency of 500~MHz and falling 
sharply at frequencies above 3~GHz due to strong synchrotron losses.

\begin{figure}
\plotone{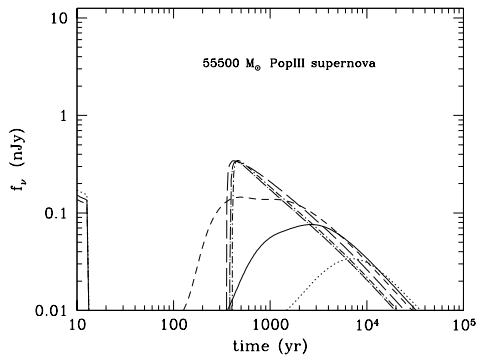} 
\caption{Radio light curves for the supermassive SN in a $3.2\times10^8$ \Ms\ halo 
at $z=15$ (observer's frame).  The curves correspond to the bands:\ 0.5 (dotted), 
1.4 (solid), 3 (short-dashed), 10 (long-dashed), 25 (dot short-dashed) and 35 (dot 
long-dashed) GHz.  }
\label{fig:radio2}
\end{figure}

The expected flux counts for the radio remnant are shown in the right panel of Fig.
\ref{fig:Mh4e7_lightcurves}, assuming one SN per halo more massive than $4 
\times10^7$ \Ms\ (the numbers would have to be reduced proportionately for fewer 
SNe per halo.) Several sources brighter than $1\,\mu{\rm Jy}$ at frequencies below 
10~GHz would be visible in a square degree field.  However, we find that the radio
signature of the explosion in the more massive \textit{Enzo} protogalaxy is much
dimmer, as we show in Fig.~\ref{fig:radio2}.  This is due to the fact that the shock
is cooler in the higher densities of the more massive halo because of the much  
higher recombination rates.  As a result, such explosions emit far less synchrotron
radiation and will not be visible even to SKA.

\begin{figure}
\plotone{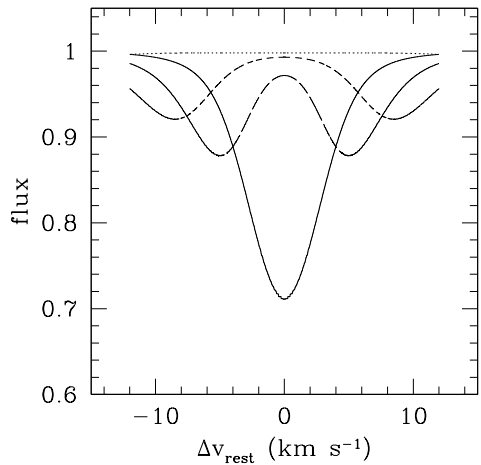} 
\caption{The 21 cm absorption signal against a bright background radio source for 
a remnant in a $4\times10^7$ \Ms\ halo at $z=15$. The curves correspond to lines 
of sight following the expansion of the remnant, with an impact parameter from the 
center of the halo of $b_\perp=110$~pc at 1.2~Myr after detonation (dotted line), 
$b_\perp=200$~pc at 4.7~Myr (short-dashed line), $b_\perp=240$~pc at 7.1~Myr 
(long-dashed line) and $b_\perp=330$~pc at 13.5~Myr (solid line).}
\label{fig:Mh4e7_21cm}
\vspace{0.1in}
\end{figure}

The long lifetimes of the remnants offer the possibility of detecting their 21 cm 
signature against a bright background radio source. While absorption from halos 
is predicted from cosmological structure formation, the signal from a given halo is 
a single absorption feature (although sometimes a doubly troughed blend) \citep{
meik11}.  As shown in Fig.~\ref{fig:Mh4e7_21cm}, the expansion of the SN 
remnant produces two distinct absorption troughs along lines of sight over a wide 
range in radii ($ \sim 100 - 300$~pc), which persist for $ \sim10^7$~yr before 
merging into a single feature. Such a unique signature would distinguish halos 
with such powerful SNe from more quiescent ones.  Assuming one such SN per 
halo, the number of such absorption features per redshift at $z > 15$ against a 
bright background radio source is $dN/dz\sim0.0001-0.001$, much higher than the
frequency found for much lower-mass SN progenitors \citep{mw12}.  It is possible 
that sufficient numbers of bright radio sources exist at these redshifts to allow the 
detection of the features.  The detection rate would be proportionately smaller
for fewer SNe per halo.

\section{Conclusion}

Supermassive Pop III SNe in dense halos briefly engulf the entire protogalaxy but 
then fall back into it in a spectacular manner, promptly enriching it to high 
metallicities.  Subsequent mergers with other halos and cold accretion flows into 
the center of the protogalaxy further mix these metals throughout its interior.  
Metals radically alter cooling in the halo in spite of the LW background, potentially 
fragmenting a large fraction of its baryons into dense clumps and igniting a brilliant 
starburst.  These bursts of early star formation would have lit up the early cosmos, 
creating large \HII\ regions \citep{wc09} and driving strong winds into the IGM 
\citep{mml99,fujita04}.  Indeed, by triggering such starbursts supermassive Pop III 
SNe may have been drivers of early reionization.  These bursts would also have 
created stellar populations that easily distinguished these special galaxies from 
others at the same redshift because their luminosities and metallicities would be 
greater than those of their more slowly evolving neighbors.   

We note that these explosions could not have been studied in our previous work
\citep{jet13a} because at early stages, when the SN remnant loses most of its 
energy, x-ray and line cooling timescales are $\sim$ 100 s, far too short to be 
tractable in 3D.  However, by the time the SN has lost most of its energy and its 
cooling times have become large, its radius in both halos is less than $\sim$ 30 
pc.  At this radius, departures from spherical symmetry in the halo would not have 
imparted significant asymmetries to the ejecta and Rayleigh-Taylor instabilities 
due to reverse shocks would not have had time to form.  It is therefore possible 
to initialize these explosions in 3D codes at later times to study their evolution in 
cosmological flows, having properly determined their energy and momentum 
losses up to that point.  Fallback, prompt chemical enrichment and starbursts in 
LW protogalaxies in such flows are now being explored in GADGET \citep{jet13b}.

The implications of primeval starbursts for the nucleosynthetic imprint of early 
SNe on the first galaxies \citep{cooke11,fb12} and on ancient, dim metal-poor 
stars \citep{bc05,fet05,Cayrel2004,Lai2008,caffau12,ren12}, remain unclear.  
Thermonuclear yields for supermassive Pop III SNe have only begun to be 
examined and may vary strongly with progenitor mass and explosion energy.  
A single massive explosion would also have enriched the entire protogalaxy to 
metallicities above those targeted by surveys of metal poor stars to date \citep{
karl08}. Such galaxies would therefore have eluded discovery in the fossil record
thus far.  Elemental yields for supermassive Pop III SNe will be the focus of 
future simulation campaigns in order to reconcile them with coming observations.

Massive fallback may also have fueled the rapid growth of 10$^4$ - 10$^6$ 
\Ms\ BHs from other supermassive clumps that formed from atomic cooling 
in the halo.  As shown in Fig.~\ref{fig:fallback}, fallback rates at the center of 
the protogalaxy can be as high as 1 \Ms\ yr$^{-1}$ for up to 10$^7$ Myr, 
perhaps driving BH accretion rates above the Eddington limit there.  Although 
these infall rates are high, they are also intermittent, so it is not yet known how 
quickly the BHs can acquire mass. How x-ray feedback from the BH regulates 
fallback is also unknown. Nevertheless, given these uncertainties it is still likely 
that these BHs, which might be the seeds of SMBHs, will have episodes of 
rapid and perhaps super-Eddington growth.  High-redshift protogalaxies with 
star formation rates much higher than those of other halos of similar mass, 
along with detections of the most energetic SNe in the universe in the NIR, 
radio and CMB, may soon mark the birthplaces of SMBHs on the sky.

\acknowledgments

DJW acknowledges support from the Baden-W\"{u}rttemberg-Stiftung by
contract research via the programme Internationale Spitzenforschung II 
(grant P- LS-SPII/18).  JLJ and JS were supported by LANL Director's 
Fellowships.  Work at LANL was done under the auspices of the National 
Nuclear Security Administration of the U.S. Department of Energy at Los 
Alamos National Laboratory under Contract No. DE-AC52-06NA25396.  
All ZEUS-MP simulations were performed on Institutional Computing 
platforms (Pinto) at LANL.

\bibliographystyle{apj}
\bibliography{refs}

\end{document}